\documentclass{article}
\usepackage[left=2cm,top=2cm,right=3cm,bottom=2cm,nohead,nofoot]{geometry}
\usepackage{amsmath,amssymb}
\begin{document}
\title{ Integrals Involving Associated Laguerre Polynomials }
\author{  M. Annamalai, M. Vasilyev,\\
email: muthiah.annamalai@mavs.uta.edu\\
Nonlinear Optics \& Nanophotonics Lab, \\
University of Texas, Arlington
}
\date{ October, 24, 2011 }
\maketitle
\section{ Introduction }
Orthogonal polynomials and their integrals are used for basis 
representations in physics problems.
Integrals involving product of Hermite polynomials like,
\begin{eqnarray}
\label{eq_hermite}
I^{a}_{m,n,p,\ldots} = \int_{-\infty}^{\infty}\left[ \exp^{-x^2/a} H_m(x) H_n(x) \ldots \right]\mathrm{d}x  \mbox{ while } (\Re(a) > 0 ),
\end{eqnarray}
was solved in \cite{ida}. A similar product of
associated Laguerre polynomials 
\begin{eqnarray}
\label{eq_lg1}
I^{a}_{(p_1,l_1),(p_2,l_2),\ldots,(p_n,l_n)} = 
\int_{0}^{\infty}\left[ \exp^{-x^2/a} L_{p_1}^{l_1}(x) L_{p_2}^{l_2}(x) \ldots  L_{p_n}^{l_n}(x) \right]\mathrm{d}x \mbox{ while } ( \Re(a) > 0, p_i \ge 0, l_i \ge 0 ),
\end{eqnarray}
does not have a closed form expression. 
In this article we present a discrete sum formula for Eq.\eqref{eq_lg1},
by using the following identities for the
associated Laguerre polynomial and an integral formula \cite{tointeg},
\begin{eqnarray}
L_{n}^{\alpha}(x) = \sum_{m=0}^{n}\left[(-1)^{m}\left(\begin{array}{l}
n + \alpha \\
n - m \\
\end{array}\right) \frac{x^m}{m!} \right],
\end{eqnarray}
\begin{eqnarray}
\int_{0}^{\infty}x^m \exp\left(-\beta x^n\right) \mathrm{d}x =
\frac{\Gamma\left[(m+1)/n\right]}{n \beta^{\left[(m+1)/n\right]}},
\mbox{while\ }\Re(\beta) > 0, \Re(m) > 0, \Re(n) > 0.
\end{eqnarray}
\section{Results}
The closed form expression of the overlap-integral is given by the nested sum,
\begin{multline}
\label{eq_lagurre}
I^{a}_{(p_1,l_1),(p_2,l_2),\ldots,(p_n,l_n)}  = \sum_{(q_1,q_2,\ldots,q_n)=0}^{(p_1,p_2,\ldots,p_n)}
\left[
\frac{(-1)^{\sum^{n}_{i=1}q_i}}{\prod_{i=1}^{n}{q_i!}}
\prod_{i=1}^{n}{ \left(\begin{array}{l}
p_i + l_i \\
p_i - q_i
\end{array} \right)}\times \frac{a^{\left( \frac{1+\sum^{n}_{i=1}q_i}{2}\right)}}{2}
 \Gamma\left({\frac{1+\sum^{n}_{i=1}q_i }{2}}\right)
\right].
\end{multline}


\begin{thebibliography}{2}
\bibitem{ida} Ida W. Busbridge, ``Some Integrals Involving Hermite Polynomials,'' Journal London Math. Soc., \textbf{23}, pp.135-141, 1948.
\bibitem{tointeg} I. S. Gradshteyn, I. M. Ryzhik, ``Table of Integrals, Series, Products'', $4^{\mathrm{th}}$ Ed, Associated Press, 1980.
\end{thebibliography}
\end{document}